\documentclass[11pt]{article}
\usepackage{amsmath}
\usepackage{amssymb}
\usepackage{graphicx}
\usepackage{color}
\usepackage{amsthm}
\usepackage{amsfonts}
\usepackage{amscd}
\usepackage{mathrsfs}
\usepackage{bm}
\usepackage{enumerate}
\usepackage{algorithmic, algorithm}

\newcommand{\D}{\displaystyle}
\newcommand{\B}{\mathbf}
\newcommand{\bx}{\mathbf{x}}

\newcommand{\ol}[1]{\overline{#1}}
\newcommand{\myref}[1]{(\ref{#1})}

\newcommand{\dt}{\Delta \theta}

\newtheorem{remark}{\textbf{Remark}}[section]

\begin{document}
\title{Sparse Time-Frequency decomposition for multiple signals with same frequencies}

\author{Thomas Y. Hou,\thanks{Applied and Comput. Math, MC 9-94, Caltech,
Pasadena, CA 91125. {\it Email: hou@cms.caltech.edu.}} \and
Zuoqiang Shi\thanks{Mathematical Sciences Center, Tsinghua University, Beijing, China, 100084. 
{\it Email: zqshi@math.tsinghua.edu.cn.}} }

\maketitle
\begin{abstract}
  In this paper, we consider multiple signals sharing same instantaneous frequencies. This kind of data is very common in scientific and engineering problems.
  To take advantage of this special structure,  
we modify our data-driven time-frequency analysis by updating the instantaneous frequencies simultaneously.  
Moreover, based on the simultaneously sparsity approximation and fast Fourier transform, some efficient algorithms is developed.
Since the information of multiple signals is used, this method is very robust to the perturbation of noise.  
And it is applicable to the general nonperiodic signals even with missing samples or outliers. 
Several synthetic and real signals are used to test this method. The performances of this method are very promising. 
\end{abstract}

\section{Introduction}
\label{sec:intro}
Data is an important bridge connecting the human being and the natural world. 
Many important information of the real world is encoded in the data. In many applications, 
frequencies of the signal are usually very useful to reveal the underlying physical mechanism.
Therefore, in the past several decades, many researchers devoted to find efficient
time frequency analysis methods and many powerful methods have been developed, including the windowed Fourier transform \cite{Mallat09}, the wavelet transform
\cite{Daub92, Mallat09}, the Wigner-Ville distribution \cite{Flandrin99}, etc. 
Recently, an adaptive time frequency analysis method, the
Empirical Mode Decomposition (EMD) method \cite{Huang98,WH09} was developed which provides an efficient adaptive method to extract frequency information 
from complicate multiscale data.
Inspired by EMD, recently several new time frequency analysis methods have been proposed 
see e.g. the synchrosqueezed wavelet transform \cite{DLW11}, the Empirical wavelet transform \cite{DZ13}, the variational mode decomposition \cite{Gilles13}.

In the last few years, inspired by the EMD method and compressive sensing \cite{CRT06a,Candes-Tao06,Dnh06},
Hou and Shi proposed a data-driven time-frequency analysis method based on the sparsest time-frequency representation \cite{HS13}. 
In this method, the signal is assumed to fit following model:
\begin{eqnarray}
\label{decomp-f}
  f(t)=\sum_{k=1}^K a_k(t)\cos\theta_k(t) ,\quad t\in [0,1], 
\end{eqnarray}
where $a_k(t),\; \theta_k(t)$ are smooth functions, $\theta_k'(t)>0,\;k=1,\cdots,M$. We assume that
$a_k(t)$ and $\theta_k'$ are "less oscillatory" than $\cos\theta_k(t)$. After some scaling, we can always make the signal lies in the time interval $[0,1]$.
So, here we assume the time span of the interval is $[0,1]$.
We borrow the terminology in EMD method and also call $a_k(t)\cos\theta_k(t)$ as the Intrinsic Mode Functions (IMFs) \cite{Huang98}. 
This model is also known as Adaptive Harmonic model (AHM) and is widely used in the time-frequency analysis literatures \cite{CM15,DLW11}.

The AHM model seems to be natural and simple In this model, the main difficulty is how to compute the decomposition. In the AHM model, the number of degrees of 
freedom is much larger than the given signal. This makes the 
possible decomposition is not unique. One essential problem is how to set up a criterion to pick up the "best" decomposition and this criterion should be 
practical in computation. Inspired by the compressive sensing, we proposed to decompose the signal by looking for the sparsest decomposition. And the sparsest 
decomposition is obtained
by solving a nonlinear optimization problem formulated as following:
 \begin{eqnarray}
\label{opt-l0}
\min_{(a_k)_{1\le k\le K}, (\theta_k)_{1\le k\le M}} K,\quad\quad \mbox{Subject to:}\quad f=\sum_{k=1}^K a_k\cos\theta_k,\;\; a_k\cos\theta_k\in \mathcal{D}.
\end{eqnarray}
where $\mathcal{D}$ is the dictionary consist of all IMFs which makes the decomposition satisfies the AHM model (see \cite{HS13} for precise definition of the dictionary).

Two kinds of algorithms were proposed to solve the optimization problem \eqref{opt-l0}. The first one is based on matching pursuit \cite{MZ93, HS13} and the other
one is based basis pursuit \cite{CDS98, HS14}. 
The convergence of the algorithm based on matching pursuit was also analyzed under the assumption of certain scale separation property \cite{HST14}.

In the previous works, we consider the case that only one signal is given. However, in many application, we can obtain many different signals and these signals 
have same frequency structure. For instance, in the monitoring of buildings, usually many sensor are put in different locations of the same building to measure the vibration. 
From these sensors,  
many measurements of the vibration are obtained. Since the signals from different sensors measure the same building, they should have same instantaneous frequencies which 
associate with the natrual frequencies of the building. If we analyze the signals from different sensors individually, 
the structure that these signals share the same instantaneous frequencies is wasted. Intuitively, by taking advantage of
this structure, we may get more robust and efficient methods. In this paper, we will propose several methods by exploiting this special frequency structure.  

In this paper, we consider multiple signals sharing same instantaneous frequencies. Naturally, we modify the adaptive harmonic model \eqref{decomp-f} to deal with 
this kind of signals.
\begin{eqnarray}
\label{AHMM}
  f^j(t)=\sum_{k=1}^K a_k^j(t)\cos\theta_k(t) ,\quad t\in [0,1],\quad j=1,\cdots,M,
\end{eqnarray}
where $M$ is the number of signals, $f^j$ is the $j$th signal. For each $j=1,\cdots,M$ fixed, it is the AHM model \eqref{decomp-f}, i.e. 
$a_k^j(t),\; \theta_k(t)$ are smooth functions, $\theta_k'(t)>0,\;k=1,\cdots,K$. We assume that
$a_k^j(t)$ and $\theta_k'$ are "less oscillatory" than $\cos\theta_k(t)$. The main feature of this model is that 
different signals $f^j$ have same phase functions $\theta_k$. We call \eqref{AHMM} Multiple Adaptive Harmonic
model (MAHM).

Inspired by our previous work on single signal, we also look for the sparsest decomposition which satisfy the MAHM model \eqref{AHMM} by solving following 
optimization problem:
 \begin{eqnarray}
\label{opt-ori}
&&\min_{(a_k^j,\theta_k)_{k=1,\cdots,K}^{j=1,\cdots,M}}\quad\quad\quad\quad\quad\quad\quad K\\
&&\mbox{Subject to:}\quad f^j(t)=\sum_{k=1}^K a_k^j(t)\cos\theta_k(t),\;\;\;a_k^j(t)\cos\theta_k(t)\in \mathcal{D},\;
\quad k=1,\cdots,K, \; j=1,\cdots,M\nonumber.
\end{eqnarray}
The dictionary $\mathcal{D}$ we used in this paper is same as that in \cite{HS13}. It can be written as
\begin{eqnarray}
\label{dic-D}
  \mathcal{D}=\left\{a\cos\theta:\; a\in V(\theta,\lambda),\;\theta'\in 
V(\theta,\lambda),\mbox{and}\; \theta'(t)\ge 0, \forall 
t\in [0,1] \right\} .
\end{eqnarray}
$V(\theta,\lambda)$ is the collection of all the functions "less oscillatory"
than $\cos\theta(t)$. In general, it is most effective to construct
$V(\theta,\lambda)$ as an overcomplete Fourier basis given below:
\begin{eqnarray}
\label{2-fold-fourier-2}
V(\theta,\lambda)=\mbox{span}\left\{1, \left(\cos\left(\frac{k\theta}{2L_\theta}\right)\right)_{1\le k\le 2\lambda L_\theta}
,\left(\sin\left(\frac{k\theta}{2L_\theta}\right)\right)_{1\le k\le 2\lambda L_\theta}\right\},
\end{eqnarray}
where $L_\theta=\lfloor\frac{\theta(1)-\theta(0)}{2\pi}\rfloor$,
$\lfloor\cdot \rfloor$ is the largest integer less than $(\cdot)$,
 and $\lambda\le 1/2$ is a parameter to control the smoothness of 
$V(\theta, \lambda)$. 
In our computations, we typically choose $\lambda =1/2$.

In the rest of the paper, we will introduce several algorithms to approximately solve above optimization problem \eqref{opt-ori}. First, we give a generic algorithm 
based on matching pursuit and nonlinear least squares in Section 2. This algorithm can be accelerated by fast Fourier transform (FFT) if the signal is periodic. This 
will be reported in Section 3. For general nonperiodic signals, in Section 4, we also develop an efficient algorithm based on group sparsity and the algorithm in Section 2.
This algorithm can be also generalized to deal with the signals with outliers or missing samples (Section 5). In Section 6, we present several numerical results 
including both synthetic and real signal to demonstrate the performance of our method. At the end, some conclusion remarks are made in Section 7.

\section{Method based on matching pursuit}
\label{sec:mp}
Following the same idea as that in \cite{HS13}, we proposed a method based on mathcing pursuit to get the sparsest decomposition, see Algorithm \ref{alg:mp}.
\begin{algorithm}
\floatname{algorithm}{Algorithm}
\caption{Method based on matching pursuit}
\label{alg:mp}
\begin{algorithmic}[1]
\REQUIRE Signals $f^j(t)$, $j=1,\cdots,M$.
\ENSURE Phase functions and the corresponding envelopes: $\theta_k,\;a_k^j,\quad k=1,\cdots,K,\;\;j=1,\cdots,M$.
\STATE Set $r^j_1=f^j(t), \; j=1,\cdots,M$ and $k=1$.
\WHILE{ $\D \max_j\left(\|r^{j}_k\|_{l^2}\right)>\epsilon_0$}
\STATE Solve the following nonlinear least-square problem:
\begin{eqnarray}
\label{opt-greedy}
&&\min_{a_k^j,\theta_k}\quad
\sum_{j=1}^M\|r^j_{k}-a^j_{k}\cos\theta_{k}\|_{l^2}^2\\
&&\mbox{Subject to:}\quad\theta'_k\ge 0,
\quad a_k^j(t)\in V(\theta_k),\;\; j=1,\cdots,M, \nonumber.
\end{eqnarray}
\STATE Update the residual
\begin{eqnarray*}
r^j_{k+1}=r^j_{k}-a_k^j(t)\cos\theta_k,\; j=1,\cdots,M.
\end{eqnarray*}
\STATE Set $k=k+1$.
\ENDWHILE
\end{algorithmic}
\end{algorithm}

To solve the nonlinear least-squares problem \eqref{opt-greedy} in Algorithm \ref{alg:mp}, we use the well-known Gauss-Newton type iteration. 
This algorithm is also very similar as that in \cite{HS13}. The only difference is the update of the phase function. In our model, 
different signals share the same phase function. So in the algorithm, we only update one common phase function by taking some average 
among different signals. 
\begin{algorithm}
\floatname{algorithm}{Algorithm}
\caption{Gauss-Newton iteration to solve the nonlinear least-squres problem}
\label{alg:gauss-newton}
\begin{algorithmic}[1]
\REQUIRE Initial guess of the phase function $\theta_k^0=\theta_0$.
\ENSURE Phase functions and the corresponding envelopes: $\theta_k,\;a_k^j,\quad j=1,\cdots,M$.
\WHILE{ $\|\theta_k^{n+1}-\theta_k^n\|_{l^2}>\epsilon_0$}
\STATE  Solve the following least-square problem for each signal $r_{k-1}^j$, $j=1,\cdots,M$:
\begin{eqnarray}
\label{opt-linear-ls-arctan}
&&\min_{a_k^{j,n+1},b_k^{j,n+1}}\quad
\|r^{j}_{k-1}-a_{k}^{j,n+1}(t)\cos\theta_{k}^{n}(t)-b_k^{j,n+1}(t)\sin\theta_k^n(t)\|_{l^2}^2 \\
&&\mbox{Subject to}\quad a_{k}^{j,n+1}(t),\;b_{k}^{j,n+1}(t)\in V(\theta_{k}^n).\nonumber
\end{eqnarray}
\STATE Calculate averaged updation of frequency:
\begin{eqnarray}
\Delta\omega_j=\frac{a_k^{j,n+1}\left(b_k^{j,n+1}\right)'-
b_k^{j,n+1}\left(a_k^{j,n+1}\right)'}{\left(a_k^{j,n+1}\right)^2+\left(b_k^{j,n+1}\right)^2},\quad 
\Delta\omega=\frac{\sum_{j=1}^M\Delta\omega_j\;\Gamma_k^{j,n+1}}{\sum_{j=1}^M\Gamma_k^{j,n+1}},
\end{eqnarray}
    
where $\Gamma_k^{j,n+1}=\left(a_k^{j,n+1}\right)^2+\left(b_k^{j,n+1}\right)^2$ and $(\cdot)'$ denote the derivative of $(\cdot)$ with respect to  $t$.
\STATE Update $\theta_k^n$
\begin{eqnarray*}
\dt=\int_0^t\Delta\omega(s)ds,\quad
\theta_k^{n+1}=\theta_k^n-\beta \dt,
\end{eqnarray*}
where $\beta\in [0,1]$ is chosen to make sure that
 $\theta_k^{n+1}$ is monotonically increasing:
\begin{eqnarray*}
\beta=\max\left\{\alpha\in [0,1]: \frac{d}{dt}\left(\theta_k^n-\alpha
\dt\right)\ge 0\right\}.
\end{eqnarray*}
\ENDWHILE

\end{algorithmic}
\end{algorithm}

By integrating Algorithm \ref{alg:mp} and \ref{alg:gauss-newton} together, we can get a complete algorithm to compute the sparsest decomposition. 
The key step and also most expensive step is to solve the least-squares problem \eqref{opt-linear-ls-arctan}. It does not need too much time to 
solve \eqref{opt-linear-ls-arctan} once, however, we have to solve this least-squares problem many times to get the final decomposition. This makes that 
the algorithm is not very practical in applications. In next two sections, we will propose some acceleration algorithms.   

\section{Periodic signals}
\label{sec:periodic}
If the signals are periodic, we can use a 
standard Fourier basis to construct 
the $V(\theta,\lambda)$ space instead of the over-complete Fourier basis given in \myref{2-fold-fourier-2}. 
\begin{eqnarray}
\label{def-Vp}
  V_p(\theta,\lambda)=\mbox{span}\left\{1, \left(\cos\left(\frac{k\theta}{L_\theta}\right)\right)_{1\le k\le \lambda L_\theta}, 
\left(\sin\left(\frac{k\theta}{L_\theta}\right)\right)_{1\le k\le \lambda L_\theta}\right\},
\end{eqnarray}
where $\lambda\le 1/2$ is a parameter to control the smoothness of
functions in $V_p(\theta,\lambda)$ and $L_\theta=(\theta(1)-\theta(0))/2\pi$ is a positive integer.

In this case, we developed an efficient algorithm based fast Fourier transform to solve the least-squares problem \eqref{opt-linear-ls-arctan} in 
\cite{HS13}. To make this paper self-contain, we also include the algorithm here. 



Suppose that the signal $r_{k-1}$ is measured over a uniform grid 
$t_j=j/N, \; j=0,\cdots, N-1$. Here, we assume that the sample points are fine enough such that $r_{k-1}$ can be
interpolated to any grid with small error. 
Let $\ol{\theta}=\frac{\theta-\theta(0)}{\theta(1)-\theta(0)}$ 
be the normalized phase function and $L_{\theta}=\frac{\theta(1)-\theta(0)}{2\pi}$ which is an integer.

The FFT-based algorithm to approximately solve \eqref{opt-linear-ls-arctan} is given below:
\begin{itemize}
\item[Step 1:] Interpolate $r_{k-1}$ from $\{t_i\}_{i=1}^N$ in the
physical space  to a uniform mesh in the
 $\theta^n_k$-coordinate to get $r_{\theta^n_k}$ and compute the Fourier transform 
$\widehat{r}_{\theta^n_k}$:
  \begin{eqnarray}
    r_{\theta^n_k,\,j}=\mbox{Interpolate}\;\left(r_{k-1},\theta^n_{k,\,j}\right),
  \end{eqnarray}
where $\theta^n_{k,\,j}, \; j=0,\cdots,N-1$ are uniformly distributed in the
 $\theta^n_k$-coordinate,i.e. $\theta^n_{k,\,j}=2\pi L_{\theta }\; j/N$. And the Fourier transform of 
$r_{\theta^n_k}$ is given as follows
  \begin{eqnarray}
    \widehat{r}_{\theta^n_k}(\omega)=\frac{1}{N}\sum_{j=1}^N r_{\theta ,\,j}
e^{-i2\pi \omega \ol{\theta}^n_{k,\,j} },\quad \omega=-N/2+1,\cdots,N/2, 
  \end{eqnarray}
where $\ol{\theta}^n_{k,\,j}=\frac{\theta^n_{k,\,j}-\theta^n_{k,\,0}}{2\pi L_{\theta^n_k}}$.

\item[Step 2:] Apply a cutoff function to the Fourier Transform of
$r_{\theta^n_k}$ to compute $a$ and $b$ on the mesh of
the $\theta^n_k$-coordinate, denoted by $a_{\theta^n_k}$ and $b_{\theta^n_k}$:
\begin{eqnarray}
a_{\theta^n_k}&=&\mathcal{F}^{-1}\left[\left(\widehat{r}_{\theta^n_k}\left(\omega+L_{\theta^n_k}\right)+
\widehat{r}_{\theta^n_k}\left(\omega-L_{\theta^n_k}\right)\right)\cdot 
\chi_\lambda\left(\omega/L_{\theta^n_k}\right)\right],\\
b_{\theta^n_k}&=&\mathcal{F}^{-1}\left[i\cdot\left(\widehat{r}_{\theta^n_k}\left(\omega+L_{\theta^n_k}\right)-
\widehat{r}_{\theta^n_k}\left(\omega-L_{\theta^n_k}\right)\right)\cdot 
\chi_\lambda \left(\omega/L_{\theta^n_k}\right)\right]
.
\end{eqnarray}
$\mathcal{F}^{-1}$ is the inverse Fourier transform defined in the $\theta $ coordinate:
\begin{eqnarray}
    \mathcal{F}^{-1}\left(\widehat{r}_{\theta^n_k}\right)=\frac{1}{N}\sum_{\omega=-N/2+1}^{N/2} \widehat{r}_{\theta^n_k}
e^{i2\pi \omega \ol{\theta}^n_{k,\,j} },\quad j=0,\cdots,N-1. 
  \end{eqnarray}
\item[Step 3:] Interpolate $a_{\theta }$ and $b_{\theta }$
from the uniform mesh $\{\theta^n_{k,\,j}\}_{j=1}^N$ in 
the $\theta^n_k$-coordinate back to 
the physical grid points $\{t_i\}_{i=1}^N$:
  \begin{eqnarray}
    a(t_i) &=&\mbox{Interpolate}\;\left(a_{\theta },t_i\right),\quad i=0,\cdots,N-1,\\
    b(t_i) &=&\mbox{Interpolate}\;\left(b_{\theta },t_i\right),\quad i=0,\cdots,N-1,.
  \end{eqnarray}
\end{itemize}

The low-pass filter $\chi_\lambda (\omega)$ in the second step is determined by the choice of $V_p(\theta,\lambda)$.

In this paper, we choose the following low-pass filter 
$\chi_\lambda(\omega )$ to define $V_p(\theta,\lambda)$:
\begin{eqnarray}
  \label{cutoff-cosine}
\chi_\lambda (\omega)=\left\{
  \begin{array}{cl}
    1+\cos(\pi \omega/\lambda) ,& -\lambda <\omega< \lambda \\
0,& \mbox{otherwise}.
  \end{array}
\right.
\end{eqnarray}


By incorporating the FFT-based solver in Algorithm \ref{alg:gauss-newton},
we get an FFT-based iterative algorithm which is summarized in Algorithm \ref{alg:fft}.
\begin{algorithm}
\floatname{algorithm}{Algorithm}
\caption{(FFT-based algorithm to solve the nonlinear least-squres problem)}
\label{alg:fft}
\begin{algorithmic}[1]
\REQUIRE Initial guess of the phase function $\theta_k^0=\theta_0,\quad \eta=0$.
\ENSURE Phase functions and the corresponding envelopes: $\theta_k,\;a_k^j,\quad j=1,\cdots,M$.
\WHILE {$\eta<\lambda$}
\WHILE{ $\|\theta_k^{n+1}-\theta_k^n\|_{l^2}>\epsilon_0$}
\STATE  Interpolate $r_{k-1}$ to a uniform mesh in the
 $\theta^n_k$-coordinate to get $r_{\theta^n_k}$ and compute the Fourier transform 
$\widehat{r}_{\theta^n_k}$.
\STATE  Apply a cutoff function to the Fourier Transform of
$r_{\theta^n_k}$ to compute $a$ and $b$ on the mesh of
the $\theta^n_k$-coordinate, denoted by $a_{\theta^n_k}$ and $b_{\theta^n_k}$.
\STATE  Interpolate $a_{\theta }$ and $b_{\theta_k^n}$ back to the uniform mesh of $t$.
\STATE Calculate averaged updation of frequency:
\begin{eqnarray}
\Delta\omega_j=\frac{a_k^{j,n+1}\left(b_k^{j,n+1}\right)'-
b_k^{j,n+1}\left(a_k^{j,n+1}\right)'}{\left(a_k^{j,n+1}\right)^2+\left(b_k^{j,n+1}\right)^2},\quad 
\Delta\omega=\frac{\sum_{j=1}^M\Delta\omega_j\;\Gamma_k^{j,n+1}}{\sum_{j=1}^M\Gamma_k^{j,n+1}},
\end{eqnarray}
    
where $\Gamma_k^{j,n+1}=\left(a_k^{j,n+1}\right)^2+\left(b_k^{j,n+1}\right)^2$ and $(\cdot)'$ denote the derivative of $(\cdot)$ with respect to  $t$.
\STATE Update $\theta_k^n$
\begin{eqnarray*}
\dt' = P_{V(\theta;\,\eta)}\left(\Delta\omega\right),\quad
\dt(t)=\int_0^t\dt'(s)ds,\quad
\theta_k^{n+1}=\theta_k^n-\beta \dt,
\end{eqnarray*}
where $\beta\in [0,1]$ is chosen to make sure that
 $\theta_k^{n+1}$ is monotonically increasing:
\begin{eqnarray*}
\beta=\max\left\{\alpha\in [0,1]: \frac{d}{dt}\left(\theta_k^n-\alpha
\dt\right)\ge 0\right\}.
\end{eqnarray*}
and $P_{V_p(\theta;\,\eta)}$ is the projection operator to the space $V_p(\theta;\,\eta)$.
\ENDWHILE
\STATE $\eta=\eta+\Delta\eta$
\ENDWHILE
\end{algorithmic}
\end{algorithm}

\begin{remark}
We remark that the projection operator $P_{V_p(\theta;\,\eta)}$ is in fact
a low-pass filter in the $\theta$-space. For non-periodic data, we apply
a mirror extension to $\dt$ before we apply the low-pass filter.

 Gauss-Newton type iteration is sensitive to the 
initial guess. In order to abate the 
dependence on the initial guess, we gradually 
increase the value of $\eta$ to improve the approximation to the 
phase function so that it converges to the correct value. The detailed explanation can be found in \cite{HS13}.
\end{remark}

\section{nonperiodic signals}
\label{sec:nonperiodic}

In most of the applications, the signals are not periodic. To apply the FFT-based algorithm in previous section, we have to do periodic extension for 
general nonperiodic signals. In this paper, the signals are assumed to satisfy the MAHM model \eqref{AHMM}. One consequence of the MAHM model is that
the phase function can be used as a coordinate and 
the signals are sparse over the Fourier basis in $\theta$ coordinate. Then, one natural way to do periodic extension is to look for the sparsest representation 
of the signal over the over-complete Fourier basis defined in $\theta$ coordinate.

More precisely, the extension is obtained by solving an $l_1$ minimization problem.
\begin{eqnarray}
\min_{\bx\in \mathbb{R}^{N_b}} \|\bx\|_{1},\quad\mbox{subject to}\quad \bm{\Phi}_\theta\cdot \bx=\mathbf{f} \label{opt-extension}
\end{eqnarray}
$\mathbf{f}$ is the sample of the signal at $t_j, j=1,\cdots, N_s$, $N_s$ is the number of sample points. $t_j$ may not be uniformly distributed, however, we assume that 
the sample points are fine enough such that $\mathbf{f}$ can be interpolated over any other grids without loss of accuracy. 
$\bm{\Phi}_\theta\in\mathbb{C}^{N_s\times N_b}$ is a matrix consisting of basis functions, $N_b$ is the number of Fourier modes. 
$\bm{\Phi}_\theta(j,k)=e^{ik \theta(t_j)/\bar{L}_\theta},\quad j=1,\cdots,N_s,\; k=-N_b/2+1,\cdots, N_b/2$. $\bar{L}_\theta$ is a positive integer determined 
by the length of the period of the 
extended signal which will be discussed later. 

To get the periodic extension of $M$ signals, one way is to solve \eqref{opt-extension} $M$ times independently. However, these $M$ signals are not independent. They share
the same phase functions. To take advantage of this structure, inspired by 
the methods for simultaneous sparsity approximation \cite{TGS06, Tropp06},
we propose to solve following optimization problem to get the periodic extension of $M$ signals simultaneously. 
\begin{eqnarray}
\min_{\mathbf{X}\in \mathbb{R}^{N_b\times M}} \|\mathbf{X}\|_{2,1},\quad\mbox{subject to}\quad \bm{\Phi}_\theta\cdot \mathbf{X}=\mathbf{F} \label{opt:group}
\end{eqnarray}
where $\mathbf{X}$ is a $N_b\times M$ matrix, $\mathbf{X}=(x_{j,k})_{j=1,\cdots,N_b, k=1,\cdots M}$, $\mathbf{F}=[\mathbf{f}^1,\cdots,\mathbf{f}^M]$ is a $N_s\times M$ matrix and
\begin{equation}
  \label{eq:norm21}
  \|\mathbf{X}\|_{2,1}=\sum_{j=1}^{N_b}\left(\sum_{k=1}^M|x_{j,k}|^2\right)^{1/2} 
\end{equation}
The problem left is how to solve the optimization problem \eqref{opt:group} efficiently. Since the sample points are fine enough, we can assume that the sample points are
uniformly distributed in $\theta$ coordinate, i.e., $\theta(t_j)=(j-1)\Delta \theta$ and $\Delta \theta=2\pi\bar{L}_\theta/N_b$. Otherwise, we could use some interpolation 
method (for instance cubic spline interpolation) to get the signals over the uniformly distributed sample points. Now, we define a matrix, 
$\bar{\bm{\Phi}}_\theta\in \mathbb{C}^{N_b\times N_b}$, 
consisting of complete Fourier basis,
\begin{equation}
  \label{eq:matrix-fourier}
  \bar{\bm{\Phi}}_\theta(j,k)=\frac{1}{\sqrt{N_b}}e^{ik (j-1)\Delta\theta/\bar{L}_\theta}, \quad j=1,\cdots,N_b,\;\; k=1,\cdots, N_b.
\end{equation}
$\bar{\bm{\Phi}}_\theta(j,k)$ represents the element of $\bar{\bm{\Phi}}_\theta$ at $j$th row and $k$th column. Using the property of the Fourier basis, we know that 
$\bar{\bm{\Phi}}_\theta$ is an orthonormal matrix which will give us lots of convenience in deriving the fast algorithm. 

For the convenience of notation, denote $\Omega$ be the set of index of $\mathbf{F}$,
\begin{equation}
  \label{eq:index}
  \Omega=\{(j,k): j=1,\cdots,N_b,\; (j-1)\Delta\theta\le \theta(1)-\theta(0),\; k=1,\cdots, N_b\}.
\end{equation}
Then, for any $(j,k)\in \Omega$, $\mathbf{F}(j,k)$ is a sample point. We also define an extension of $\mathbf{F}$ by zero padding,
\begin{equation}
  \label{eq:matrix-signal}
  \bar{\mathbf{F}}(j,k)=\left\{\begin{array}{cc} \mathbf{F}(j,k),& (j,k)\in \Omega,\\
0,& \text{otherwise}.
\end{array}\right.
\end{equation}
where $j,k=1,\cdots,N_b$.

First, we remove the constraint in \eqref{opt:group} by introducing a penalty term,
\begin{eqnarray}
\min_{\bx\in \mathbb{R}^{N_b}} \|\mathbf{X}\|_{2,1}+\frac{\mu}{2} \|\bm{\Phi}_\theta\cdot \mathbf{X}-\mathbf{F}\|_2^2 \label{opt:group-penalty}
\end{eqnarray}
Here $\mu>0$ is a parameter of the penalty. We will let $\mu$ goes to infinity later. Then the solution of \eqref{opt:group-penalty} will converge to the solution of 
\eqref{opt:group}. 

Let
$\bar{\mathbf{Y}}=\bar{\bm{\Phi}}_\theta\cdot \mathbf{X}-\bar{\mathbf{F}}$, and $\mathbf{Y}=\bar{\mathbf{Y}}|_\Omega
=\bm{\Phi}_\theta\cdot \mathbf{X}-\mathbf{F}$. Using the Augmented Lagrange Multiplier method (ALM) to solve the unconstrained optimization problem \eqref{opt:group-penalty},
we get following algorithm. Let $\mathbf{Q}^0=\mathbf{0}$, and repeat following two steps until converge
\begin{itemize}
\item $\D(\mathbf{X}^{k+1},\bar{\mathbf{Y}}^{k+1})=\arg\min_{\mathbf{X},\bar{\mathbf{Y}}\in \mathbb{R}^{N_b\times M}}\|\mathbf{X}\|_{2,1}+\frac{\mu}{2} \|\mathbf{Y}\|_2^2+\frac{\gamma}{2}
\|\bar{\mathbf{Y}}-\bar{\bm{\Phi}}_\theta\cdot \mathbf{X}+\bar{\mathbf{F}}+\mathbf{Q}^k/\gamma\|_2^2$
\item $\mathbf{Q}^{k+1}=\mathbf{Q}^{k}+\gamma\cdot(\bar{\mathbf{Y}}^{k+1}-\bar{\bm{\Phi}}_\theta\cdot \mathbf{X}^{k+1}+\bar{\mathbf{F}})$
\end{itemize}
Here $\gamma>0$ is a parameter.
In above ALM iteration, the main computational load is to solve the optimization problem in the first step. Notice that the objective functional depends on two terms, 
$\mathbf{X}$ and $\bar{\mathbf{Y}}$. It is natural to minimize the functional alternately by optimizing one of these two terms when the other one is fixed like that in 
the split Bregman iteration \cite{GO09}.  
Using this idea, we get following iterative algorithm to solve \eqref{opt:group-penalty}.

Let  $\mathbf{Q}^0=0$, $\bar{\mathbf{Y}}^0=0$ and repeat
\begin{itemize}
\item $\D\mathbf{X}^{k+1}=\arg\min_{\mathbf{X}\in \mathbb{R}^{N_b\times M}}\|\mathbf{X}\|_{2,1}+\frac{\gamma}{2}
\|\bar{\mathbf{Y}}^k-\bar{\bm{\Phi}}_\theta\cdot \mathbf{X}+\bar{\mathbf{F}}+\mathbf{Q}^k/\gamma\|_2^2$
\item $\D\bar{\mathbf{Y}}^{k+1}=\arg\min_{\bar{\mathbf{Y}}\in \mathbb{R}^{N_b\times M}}\frac{\mu}{2} \|\mathbf{Y}\|_2^2+\frac{\gamma}{2}
\|\bar{\mathbf{Y}}-\bar{\bm{\Phi}}_\theta\cdot \mathbf{X}^{k+1}+\bar{\mathbf{F}}+\mathbf{Q}^k/\gamma\|_2^2$
\item $\mathbf{Q}^{k+1}=\mathbf{Q}^{k}+\gamma\cdot(\bar{\mathbf{Y}}^{k+1}-\bar{\bm{\Phi}}_\theta\cdot \mathbf{X}^{k+1}+\bar{\mathbf{F}})$
\end{itemize}
The good news is that the optimization problems in the first and second step can be solved explicitly and we can use fast Fourier transform to accelerate the computation. 
First, let us see the first optimization problem. Using the fact that $\bar{\bm{\Phi}}_\theta$ is an orthonormal matrix, we have
\begin{align}
  \label{eq:threshold_x}
  \mathbf{X}^{k+1}=&\arg\min_{\mathbf{X}\in \mathbb{R}^{N_b\times M}}\|\mathbf{X}\|_{2,1}+\frac{\gamma}{2}
\|\mathbf{X}-\bar{\bm{\Phi}}_\theta^{*}\cdot\left(\bar{\mathbf{Y}}^k+\bar{\mathbf{F}}+\mathbf{Q}^k/\gamma\right)\|_2^2\nonumber\\
=&\mathcal{S}_\gamma\left(\bar{\bm{\Phi}}_\theta^{*}\cdot(\bar{\mathbf{Y}}^k+\bar{\mathbf{F}}+\mathbf{Q}^k/\gamma)\right)
\end{align}
where $\bar{\bm{\Phi}}_\theta^{*}$ is the conjugate transpose of $\bar{\bm{\Phi}}_\theta$. Here $\mathcal{S}_\gamma$ is a shrinkage operator which is defined as following.
For any $\mathbf{v}\in \mathbb{C}^{M}$
\begin{equation}
  \label{eq:threshold-matrix}
  \mathcal{S}_\gamma(\mathbf{v})=\left\{\begin{array}{cc} \frac{\|\mathbf{v}\|_2-\gamma}{\|\mathbf{v}\|_2}\,\mathbf{v},& \|\mathbf{v}\|_2>\gamma,\\
0,&\|\mathbf{v}\|_2\le \gamma.
\end{array}\right.
\end{equation}
If $\mathbf{V}$ is a matrix and $\mathbf{v}_j$ is its $j$th row, then $S_\gamma(V)$ is a matrix of the same size as $\mathbf{V}$ and the $j$th row is defined to be 
$S_\gamma(\mathbf{v}_j)$. Notice that $\bar{\bm{\Phi}}_\theta$ is the matrix consisting of Fourier basis in $\theta$ coordinate, so the matrix-matrix multiplication can be evaluated 
by applying discrete Fourier transform of each column of the second matrix which can be accelerated by FFT.

In the second optimization problem, recall that $\mu>0$ is a penalty parameter in \eqref{opt:group-penalty} which is the larger the better. Now, let $\mu$ goes to 
infinity, then we can solve 
\begin{equation}
  \label{eq:Z}
  \bar{\mathbf{Y}}^{k+1}(j,k)=\left\{\begin{array}{cc}\left(\bar{\bm{\Phi}}_\theta\cdot \mathbf{X}^{k+1}-\bar{\mathbf{F}}-\mathbf{Q}^k/\gamma\right)(j,k),& (j,k)\in \Omega,\\
0,& \text{otherwise}.
\end{array}\right.
\end{equation}
Summarizing above derivation, we get a fast algorithm to solve \eqref{opt:group}, see Algorithm \ref{alg:group-sparsity}.
\begin{algorithm}
\floatname{algorithm}{Algorithm}
\caption{(Fourier extension by group sparsity)}
\label{alg:group-sparsity}
\begin{algorithmic}[1]
\REQUIRE $\mathbf{Q}^0=0,\;\; \bar{\mathbf{Y}}^{0}=0$.
\ENSURE Fourier coefficients on over-complete Fourier basis $\mathbf{X}$.
\WHILE{ $\D \|\mathbf{Q}^{k}-\mathbf{Q}^{k-1}\|_2>\epsilon_0$}
\STATE $\mathbf{X}^{k+1}=\mathcal{S}_{\gamma^{-1}}\left(\mathcal{F}_c(\bar{\mathbf{Y}}^k+\bar{\mathbf{F}}+\mathbf{Q}^k/\gamma)\right)$,
where $\mathcal{F}_c$ means apply Fourier transform on each column.
\STATE $\bar{\mathbf{Y}}^{k+1}(j,k)=\left\{\begin{array}{cc}\left(\bar{\bm{\Phi}}_\theta\cdot \mathbf{X}^{k+1}-\bar{\mathbf{F}}-\mathbf{Q}^k/\gamma\right)(j,k),& (j,k)\in \Omega,\\
0,& \text{otherwise}.
\end{array}\right.$
\STATE  $\mathbf{Q}^{k+1}=\mathbf{Q}^{k}+\gamma\cdot(\bar{\mathbf{Y}}^{k+1}-\bar{\bm{\Phi}}_\theta\cdot \mathbf{X}^{k+1}+\bar{\mathbf{F}})$
\ENDWHILE
\STATE $\mathbf{X}=\mathbf{X}^{k+1}$.
\end{algorithmic}
\end{algorithm}
By combining Algorithm \ref{alg:group-sparsity} and the algorithm in previous section, we can get a complete algorithm to deal with nonperiodic data. 
Before giving the complete algorithm, there is still one issue we need to address. 
In the derivation of Algorithm \ref{alg:group-sparsity}, we assume that the sample points are
uniformly distributed in $\theta$ coordinate with given $\theta$. For general signal which may not be uniformly sampled in $\theta$ coordinate, we need to 
do interpolation before applying Algorithm \ref{alg:group-sparsity}.

In order to do the interpolation, first, we need to give the interpolation points which are uniformly distributed in $\theta$ coordinate. In this paper, 
the interpolation points are chosen to be $\theta(0)+(j-1)\Delta \theta,\;\; j=1,\cdots,N_b$, $\Delta \theta=2\pi\bar{L}_\theta/N_b$. We set 
$\bar{L}_\theta= 2L_\theta$ and $L_\theta=\lfloor\frac{\theta(1)-\theta(0)}{2\pi}\rfloor$. This choice is corresponding to the 2-fold
over-complete Fourier basis used in \eqref{2-fold-fourier-2}. $N_b=2N_s$ where $N_s$ is the number of sample points of the original signal. 
The original signals are interpolated over the interpolation points by cubic spline interpolation.

Now, we have an algorithm for nonperiodic signal summarized in Algorithm \ref{alg:fft-extension}.
\begin{algorithm}
\floatname{algorithm}{Algorithm}
\caption{(Algorithm for nonperiodic signal)}
\label{alg:fft-extension}
\begin{algorithmic}[1]
\REQUIRE Initial guess of the phase function $\theta_k^0=\theta_0,\quad \eta=0$.
\ENSURE Phase functions and the corresponding envelopes: $\theta_k,\;a_k^j,\quad j=1,\cdots,M$.
\WHILE {$\eta<\lambda$}
\WHILE{ $\|\theta_k^{n+1}-\theta_k^n\|_{l^2}>\epsilon_0$}
\STATE  Interpolate $r_{k-1}$ to a uniform mesh in the
 $\theta^n_k$-coordinate to get $r_{\theta^n_k}$.
\STATE Using Algorithm \ref{alg:group-sparsity} to get the Fourier coefficients of $r_{\theta^n_k}$ on the over-complete Fourier basis. 
\STATE  Apply a cutoff function to the Fourier Transform of
$r_{\theta^n_k}$ to compute $a$ and $b$ on the mesh of
the $\theta^n_k$-coordinate, denoted by $a_{\theta^n_k}$ and $b_{\theta^n_k}$.
\STATE  Interpolate $a_{\theta }$ and $b_{\theta_k^n}$ back to the uniform mesh of $t$.
\STATE Calculate averaged update of frequency:
\begin{eqnarray}
\Delta\omega_j=\frac{a_k^{j,n+1}\left(b_k^{j,n+1}\right)'-
b_k^{j,n+1}\left(a_k^{j,n+1}\right)'}{\left(a_k^{j,n+1}\right)^2+\left(b_k^{j,n+1}\right)^2},\quad 
\Delta\omega=\frac{\sum_{j=1}^M\Delta\omega_j\;\Gamma_k^{j,n+1}}{\sum_{j=1}^M\Gamma_k^{j,n+1}},
\end{eqnarray}
    
where $\Gamma_k^{j,n+1}=\left(a_k^{j,n+1}\right)^2+\left(b_k^{j,n+1}\right)^2$ and $(\cdot)'$ denote the derivative of $(\cdot)$ with respect to  $t$.
\STATE Update $\theta_k^n$
\begin{eqnarray*}
\dt' = P_{V(\theta;\,\eta)}\left(\Delta\omega\right),\quad
\dt(t)=\int_0^t\dt'(s)ds,\quad
\theta_k^{n+1}=\theta_k^n-\beta \dt,
\end{eqnarray*}
where $\beta\in [0,1]$ is chosen to make sure that
 $\theta_k^{n+1}$ is monotonically increasing:
\begin{eqnarray*}
\beta=\max\left\{\alpha\in [0,1]: \frac{d}{dt}\left(\theta_k^n-\alpha
\dt\right)\ge 0\right\}.
\end{eqnarray*}
and $P_{V_p(\theta;\,\eta)}$ is the projection operator to the space $V_p(\theta;\,\eta)$.
\ENDWHILE
\STATE $\eta=\eta+\Delta\eta$
\ENDWHILE
\end{algorithmic}
\end{algorithm}

\section{Signals with outliers or missing samples}
\label{sec:outlier}
To deal with the signal with outliers,  we need to enlarge the dictionary to include all the impulses

In this case, the optimization problem is formulated in the
following way,
 \begin{eqnarray}
  \label{opt-outlier}
&&  \min_{\B{X}\in \mathbb{C}^{N_b\times M}\atop\B{Z}\in \mathbb{R}^{N_s\times M}}  \|\B{X}\|_{2,1}+\|\B{Z}\|_1,
\quad\quad\mbox{subject to:}\quad \bm{\Phi}_{\theta}\cdot \B{X}+\B{Z}=\B{F}.
\end{eqnarray}
where $\bm{\Phi}_\theta$ and $\mathbf{F}$ are same as those in \eqref{opt:group}

Following the similar derivation in Section \ref{sec:nonperiodic}, we obtain Algorithm \ref{alg:outlier} to solve above optimization problem \eqref{opt-outlier}.
\begin{algorithm}
\floatname{algorithm}{Algorithm}
\caption{(Algorithm for signals with outliers)}
\label{alg:outlier}
\begin{algorithmic}[1]
\REQUIRE $\mathbf{Q}^0=0,\;\; \bar{\mathbf{Y}}^{0}=0$.
\ENSURE Fourier coefficients on over-complete Fourier basis $\mathbf{X}$, estimate of outliers $\mathbf{Z}$.
\WHILE{ $\D \|\mathbf{Q}^{k}-\mathbf{Q}^{k-1}\|_2>\epsilon_0$}
\STATE $\mathbf{X}^{k+1}=\mathcal{S}_{\gamma^{-1}}\left(\mathcal{F}_c(\bar{\mathbf{Y}}^k-\bar{\mathbf{Z}}^k+\bar{\mathbf{F}}+\mathbf{Q}^k/\gamma)\right)$,
where $\mathcal{F}_c$ means apply Fourier transform on each column.
\STATE $\bar{\mathbf{Y}}^{k+1}(j,k)=\left\{\begin{array}{cc}\left(\bar{\bm{\Phi}}_\theta\cdot \mathbf{X}^{k+1}+\bar{\mathbf{Z}}^k
-\bar{\mathbf{F}}-\mathbf{Q}^k/\gamma\right)(j,k),& (j,k)\in \Omega,\\
0,& \text{otherwise}.
\end{array}\right.$
\STATE $\bar{\mathbf{Z}}^{k+1}=\mathcal{T}_{\gamma^{-1}}\left(\bar{\mathbf{Y}}^{k+1}+\bar{\mathbf{F}}+\mathbf{Q}^k/\gamma-\bar{\bm{\Phi}}_\theta^{*}\cdot \mathbf{X}\right)$,
where $\mathcal{T}_\gamma$ is a shrinkage operator. 

For any $x\in \mathbb{R}$,
\begin{equation}
  \mathcal{T}_\gamma(x)=\left\{\begin{array}{cc} \frac{|x|-\gamma}{|x|}\,x,& |x|>\gamma,\\
0,&|x|\le \gamma.
\end{array}\right.\nonumber
\end{equation}
For any matrix $\mathbf{A}=(a_{jk})$, $\mathcal{T}_\gamma(\mathbf{A})=(\mathcal{T}_\gamma(a_{jk}))$.
\STATE  $\mathbf{Q}^{k+1}=\mathbf{Q}^{k}+\gamma\cdot(\bar{\mathbf{Z}}^{k+1}-\bar{\bm{\Phi}}_\theta\cdot \mathbf{X}^{k+1}+\bar{\mathbf{F}})$
\ENDWHILE
\STATE $\mathbf{X}=\mathbf{X}^{k+1}$, $\mathbf{Z}=\bar{\mathbf{Z}}^{k+1}|_\Omega$, $\Omega$ is defined in \eqref{eq:index}.
\end{algorithmic}
\end{algorithm}
By using the same interpolation procedure in Section \ref{sec:nonperiodic}, we can integrate Algorithm \ref{alg:outlier} with the algorithm in Section \ref{sec:periodic}
to get the method to deal with signals with outliers.

For the signals with missing samples, first, we assign the value of the signal as the average of the signal at the locations where the sample was lost
 and then treat the missing sample as the outliers. 

\section{Numerical results}
\label{sec:numerics}
In this section, we use two numerical examples, one is synthetic and one is real data, to demonstrate the performance of the method proposed in this paper.
The first example is a simple synthetic signal which is used to test the robustness of the method to the perturbation of white noise. 

\noindent
{\bf Example 1: Synthetic signal with white noise}

In this example, the signals are generated as following:
\begin{eqnarray}
\label{data_chirp}
  f(t)=\cos(40\pi(t+1)^2), \quad f^j(t)=f(t)+5X^j(t),\quad j=1,\cdots,10,\; t\in[0,1].
\end{eqnarray}
where $X^j(t), \;j=1,\cdots,10$ are independent white noise with standard deviation $\sigma^2=1$. The number of sample points is 512 and the sample points are
uniformly distributed over $[0,1]$.

\begin{figure}

    \begin{center}
      \includegraphics[width=0.45\textwidth]{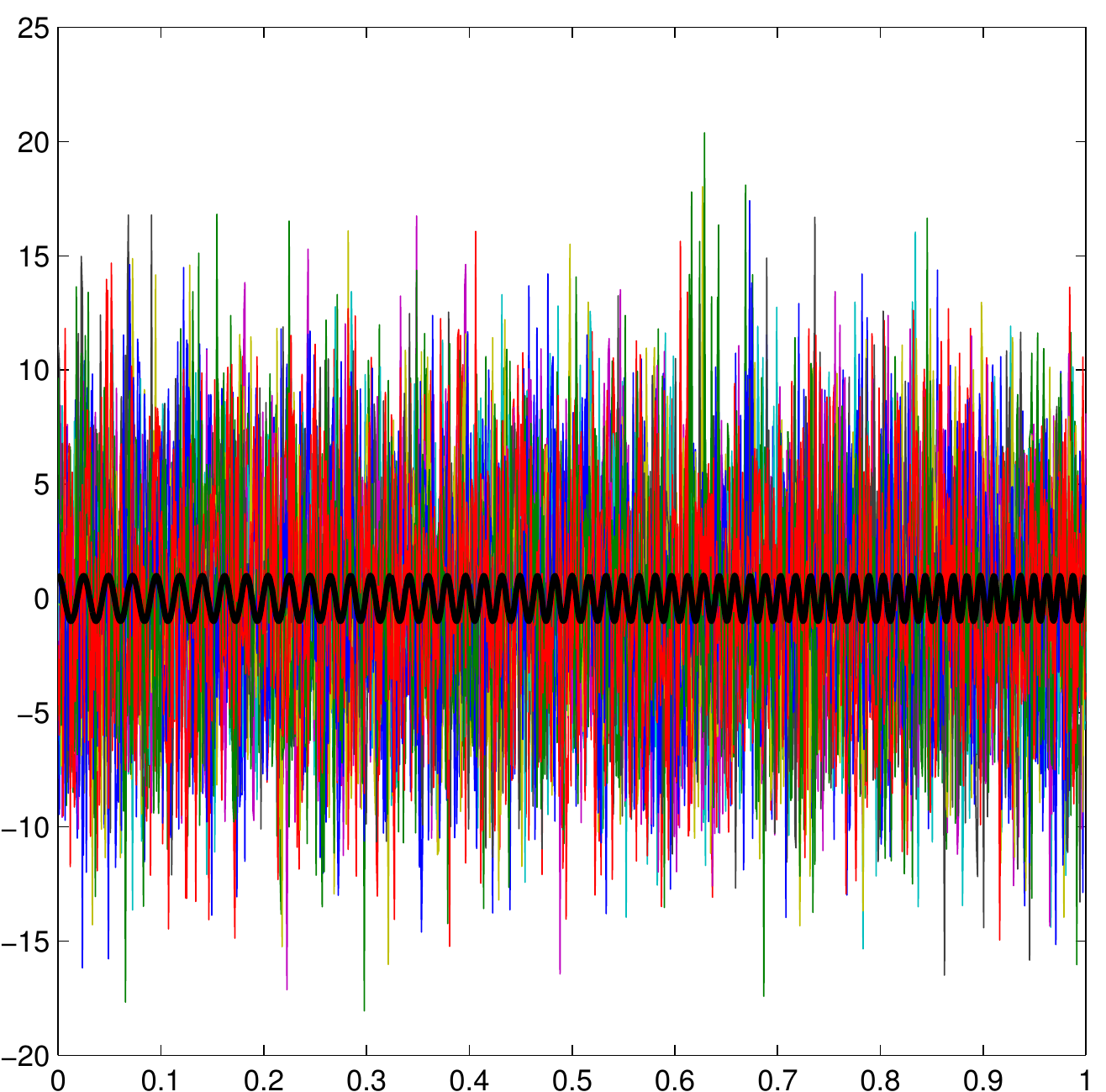}
 \includegraphics[width=0.45\textwidth]{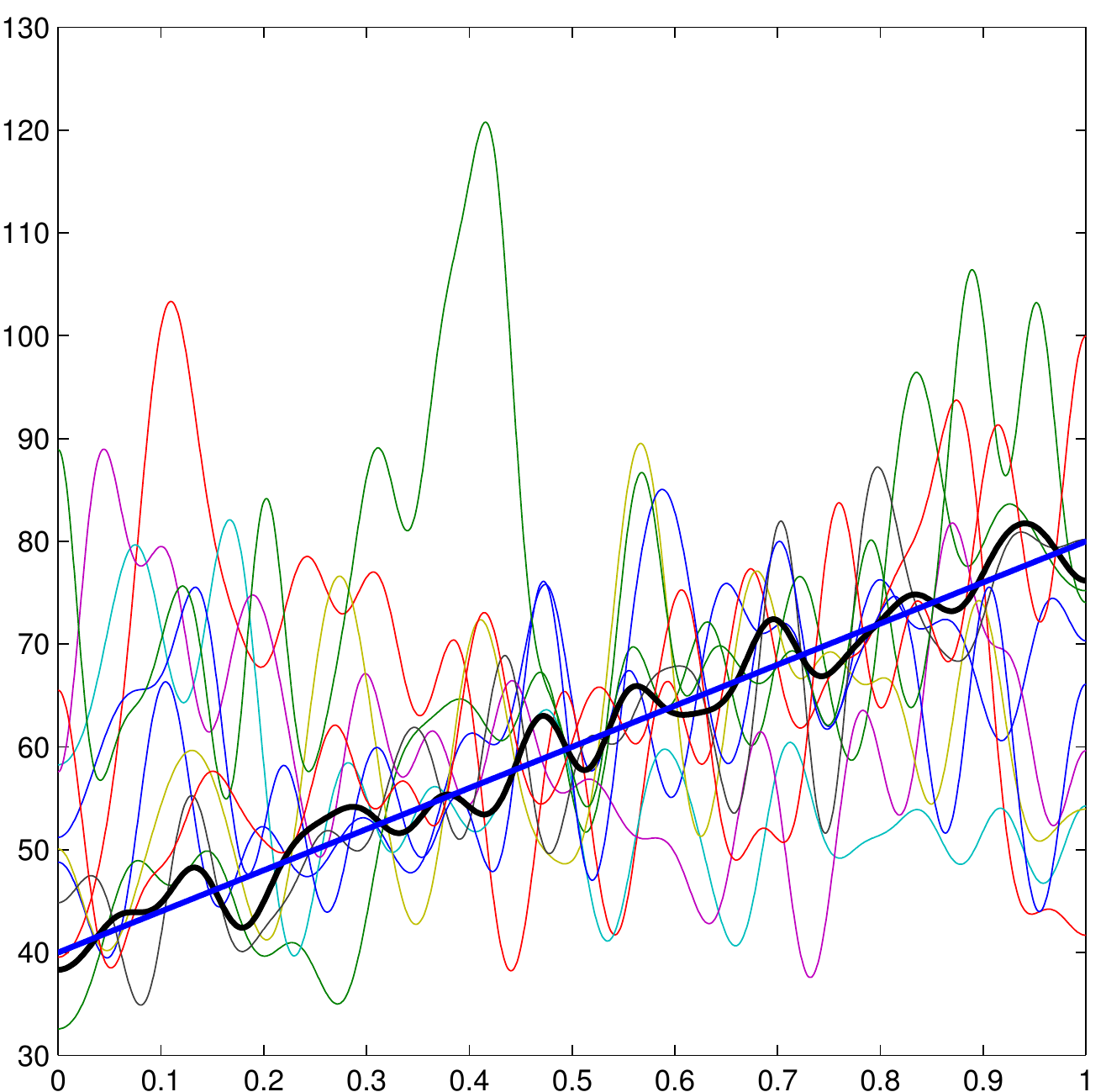}
     \end{center}
    \caption{ \label{fig:ex1} Left: 10 measurements generated by $f(t)+5X(t)$. Bold black: signal without noise;
thin lines: 10 measurements with larger noise; Right: instantaneous frequency recoverd from signal with 10 different measurements 
with larger noise.
Bold blue: exact frequency; bold black: recovered frequency from 10 signals; thin lines: frequency obtained from 10 signals
separately.}
\end{figure}

The original signals are plotted in the left figure of Fig. \ref{fig:ex1}. As we can see that the noise is so large such that we can not see any pattern of the 
original clean signal. The recovered instantaneous frequency is shown in the right figure of Fig. \ref{fig:ex1}. If we recover the frequency from each signal separately, 
since the noise is too large, the frequencies are totally wrong. However, if we use the structure that 10 signals have same instantaneous frequency,
the frequency recovered is much better. 



\noindent
{\bf Example 2: Cable tension estimate}\footnote{The authors would like to thank Prof. Yuequan Bao
of the School of Civil Engineering of Harbin Institute of Technology for the experimental data and for permission to use it in this paper.}

This example is a real problem about the tension estimate of the cables in bridge. Before demonstrating the numerical results, we give some backgrounds first.
  
For large span bridges, such as cable-stayed bridge and suspension bridge, the cables are a crucial element for overall structural safety. 
Due to the moving vehicles and other environmental effects, the cable tension forces vary over time. This variation in cable tension forces 
may cause fatigue damage. Therefore, estimation of the time-varying cable tension forces is important for the maintenance and safety assessment of cable-based bridges.

One often used method to estimate the cable tension force is based on the instantaneous frequency estimate of the cable vibration signal. 
According to the flat taut string theory that neglects both sag-extensibility and bending
stiffness, cable tension force, $F$, can be calculated by
\begin{equation}
\label{eqn:force}
  F(t)=4mL^2\left(\frac{\omega_n(t)}{2\pi n}\right)^2
\end{equation}
where $\omega_n(t)$ is the time-varying $n$th natural frequency in $radius/s$ and $m$, $L$ are mass density and
length of cable. 

Also from the flat taut string theory, an important and useful feature of the vibriations of the cable is that the natural frequencies of the 
higher modes are integer multiples of the fundamental
frequency, that is $\omega_n(t)=n\omega_1(t)$.
This feature means that we can combine the information of different modes together to recover the instantaneous frequency. Then the method developed 
in this paper for multiple signals applies after some minor modifications.  

\begin{figure}
    \begin{center}
      \includegraphics[width=0.45\textwidth]{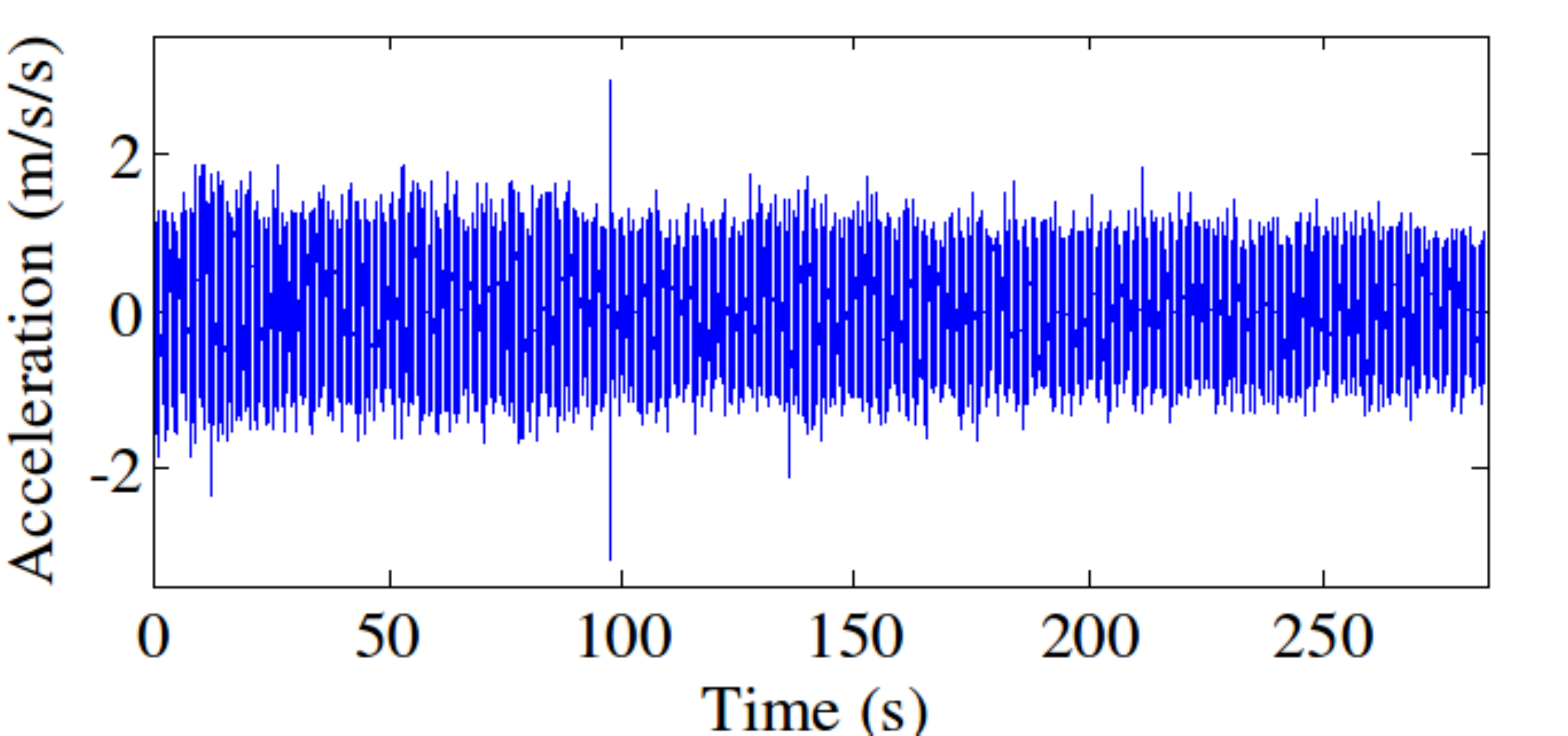}
\includegraphics[width=0.45\textwidth]{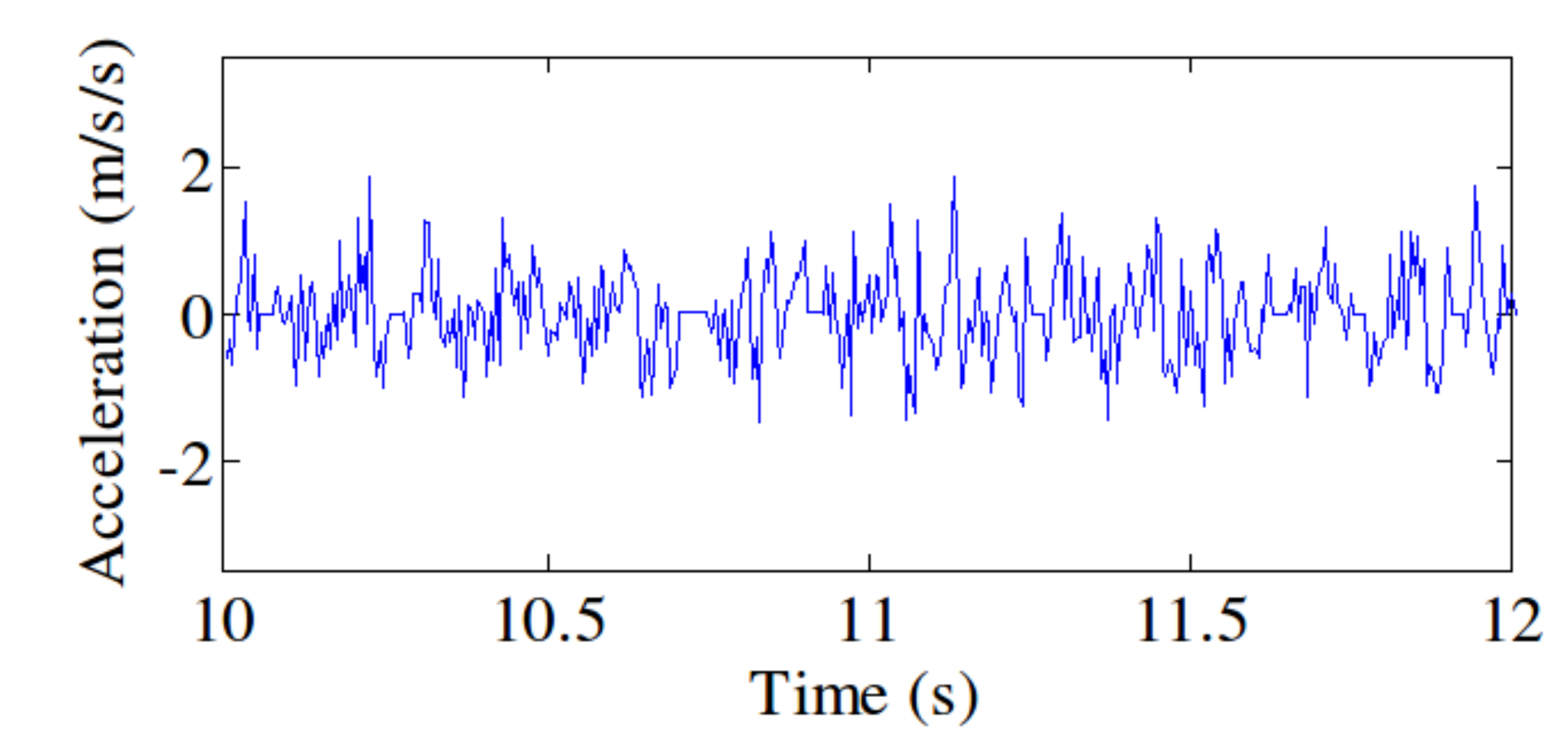}
     \end{center}
    \caption{ \label{data_cable} Left: vibration signal of the cable; Right: zoom in of the signal to demonstrate the missing samples (flat segments).}
\end{figure}

\begin{figure}

    \begin{center}
      \includegraphics[width=0.8\textwidth]{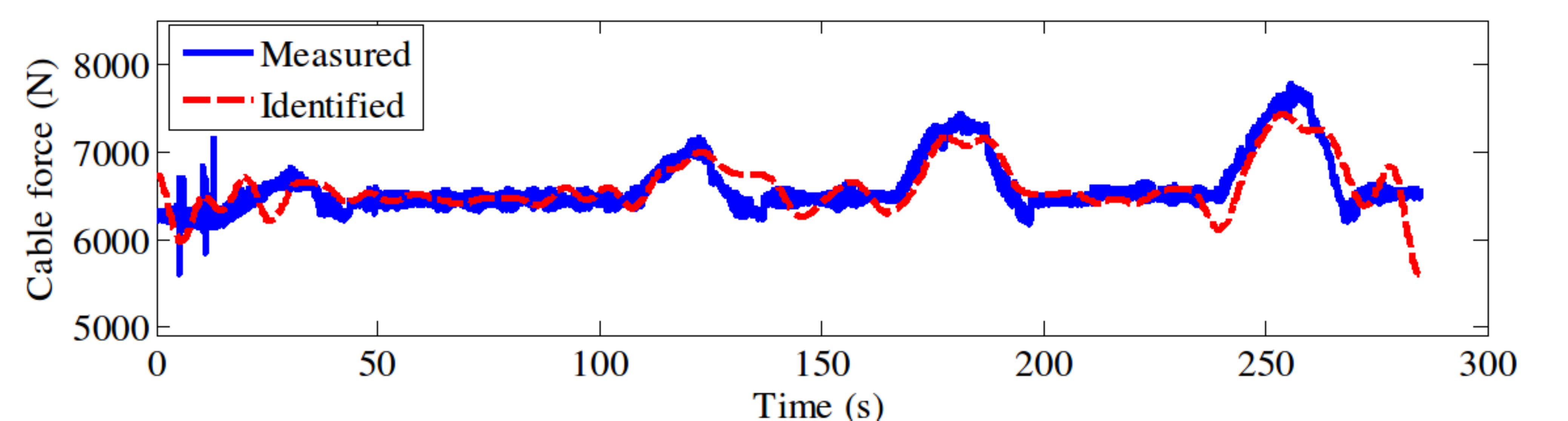}
\includegraphics[width=0.8\textwidth]{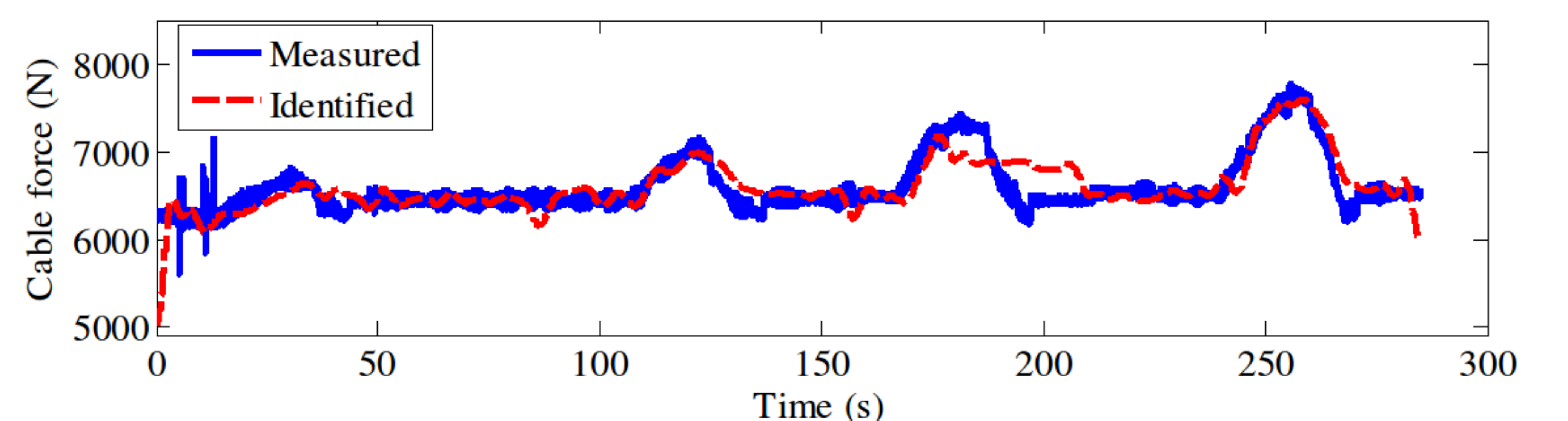}
\includegraphics[width=0.8\textwidth]{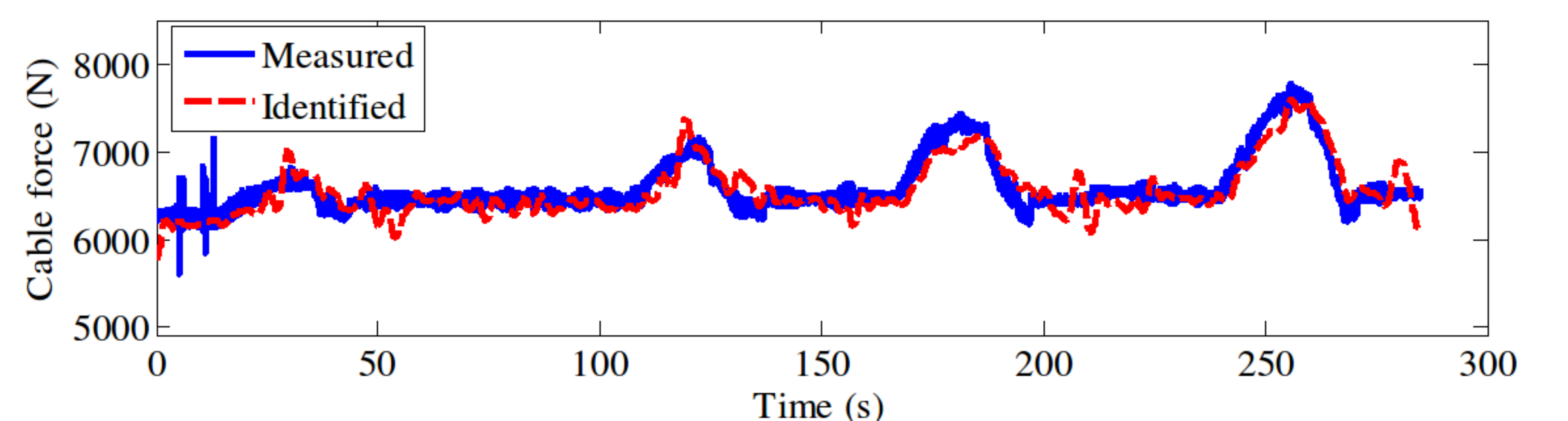}
\includegraphics[width=0.8\textwidth]{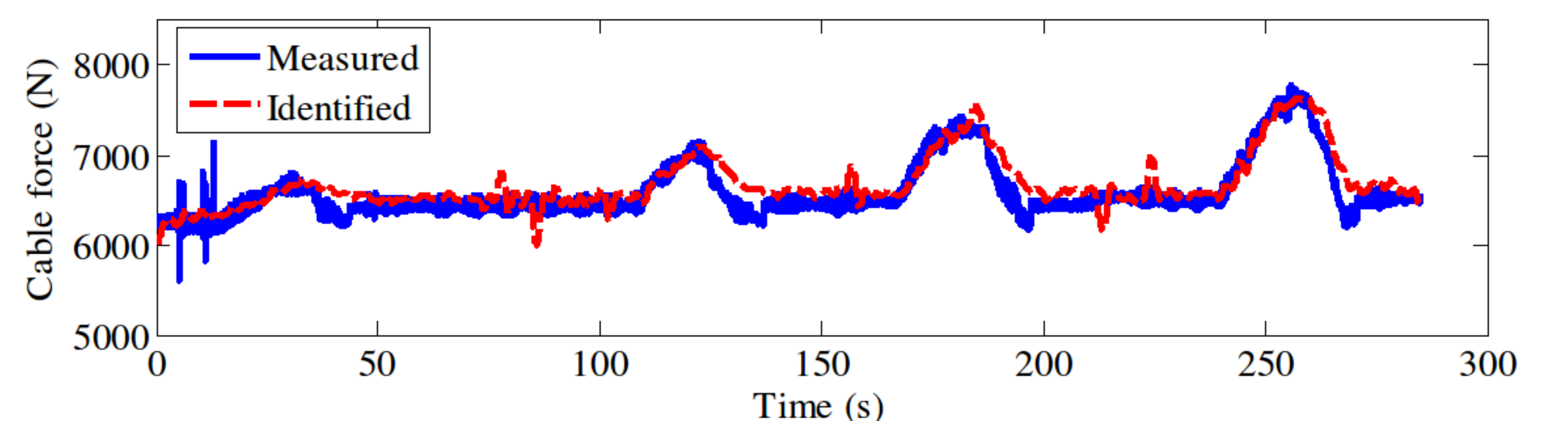}
\includegraphics[width=0.8\textwidth]{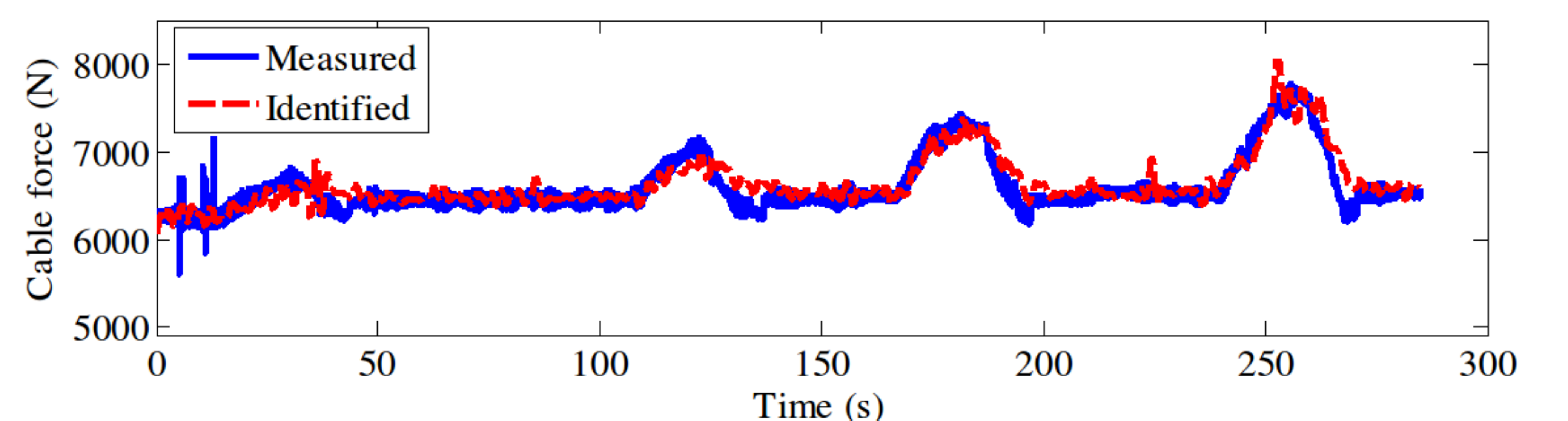}
     \end{center}
    \caption{ \label{force_single} Cable force estimated by single natural mode 1-5 from up to bottom.}
\end{figure}

\begin{figure}

    \begin{center}
      \includegraphics[width=0.8\textwidth]{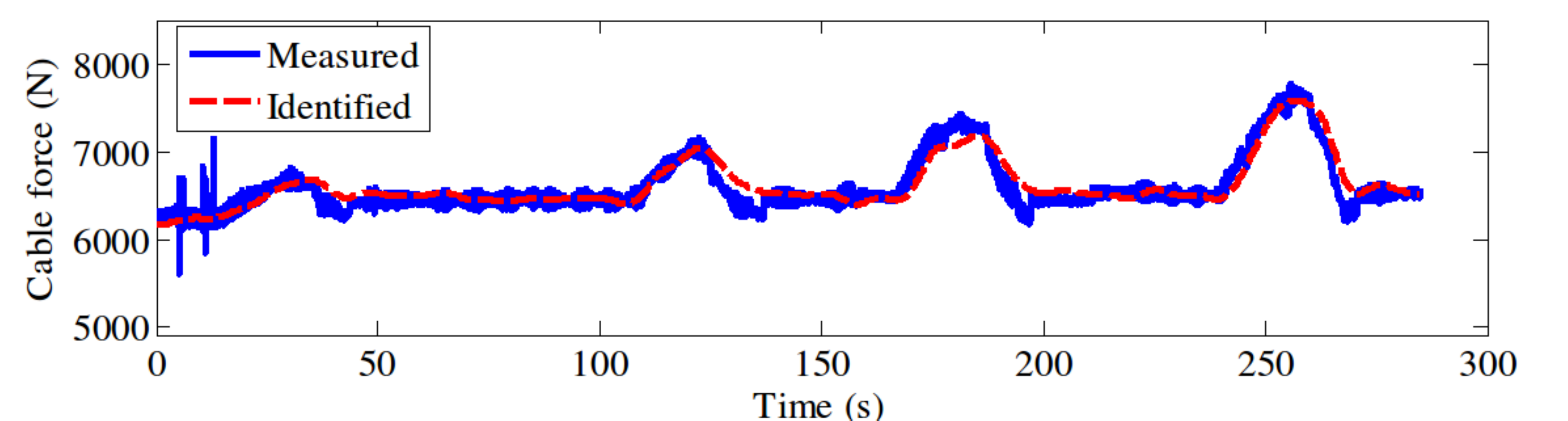}
     \end{center}
\caption{ \label{force_multi} Cable force estimated by using 1-5 natural modes together.}    
\end{figure}

The original experimental signal is given in Fig. \ref{data_cable}. Obviously, the signal has some outliers. In the original signal, about 10\% of the samples
were lost which is demonstrated very clearly in the zoom in picture in Fig. \ref{data_cable}. The tension force estimated by our method is given in Fig. \ref{force_single} and 
\ref{force_multi}. The tension forces obtained from different modes are shown in Fig. \ref{force_single}. If only one mode is used, the estimation of the force is not 
very accurate. There are many oscillations although the overall picture match with the measured force. If we use 1-5 modes together to estimate the tension force, 
the result is much better, Fig. \ref{force_multi}. 

\section{Conclusion}
\label{sec:conclusion}
In this paper, we consider multiple signals and these signals share the same instantaneous frequencies. This kind of signals emerge in many scientific and 
engineering problems. By exploiting the structure of the common frequencies, we develop some algorithms to find the sparsest time-frequency decomposition.
These algorithms can be accelerated by FFT, so they are very efficient. The algorithms are also very robust to noise, since the information of multiple 
signals are used simultaneously.

\vspace{0.2in}
\noindent
{\bf Acknowledgments.}
This work was supported by NSF FRG Grant DMS-1159138, DMS-1318377, an AFOSR MURI Grant FA9550-09-1-0613 and a DOE grant DE-FG02-06ER25727. 
The research of Dr. Z. Shi was supported by a NSFC Grant 11201257.


\end{document}